\documentclass[fleqn,usenatbib]{mnras}

\usepackage{newtxtext,newtxmath}
\usepackage{graphicx}	

\usepackage[T1]{fontenc}
\usepackage{ae,aecompl}

\usepackage{amsmath}	
\usepackage{amssymb}	

\usepackage{layout}

\newcommand{\Omegam}{\Omega_{\rm m}}

\newcommand{\Omegab}{\Omega_{\rm b}}
\newcommand{\ns}{n_{\rm s}}
\newcommand{\Vlens}{V_{\rm lens}}
\newcommand\Mpch{{\rm\; Mpc}/h}

\newcommand{\Rnl}{R_{\rm NL}}
\newcommand{\Mnl}{M_{\rm NL}}
\newcommand{\rhobgd}{\rho_{\rm bgd}}
\newcommand{\dc}{\delta_{\rm c}} 

\renewcommand{\eprint}[1]{\href{https://arxiv.org/abs/#1}{arXiv:#1}}

\title[Faster Nonlinear Mass]{Accelerating Computation
of the Nonlinear Mass by an Order of Magnitude}
\author[Krolewski \& Slepian]{
Alex Krolewski,$^{1,2}$\thanks{E-mail: krolewski@berkeley.edu (AK)} \&
Zachary Slepian$^{3,1}$\thanks{E-mail: zslepian@ufl.edu (ZS)}
\\
$^{1}$Lawrence Berkeley National Laboratory, 1 Cyclotron Road, Berkeley, CA 94720, USA\\
$^{2}$Berkeley Center for Cosmological Physics, University of California, Berkeley, Berkeley, CA 94720, USA\\
$^{3}$Department of Astronomy, University of Florida, 211 Bryant Space Science Center, Gainesville, FL 32611, USA}

\pubyear{2019}

\begin{document}
\label{firstpage}
\pagerange{\pageref{firstpage}--\pageref{lastpage}}
\maketitle







\begin{abstract}

\noindent The nonlinear mass is a characteristic scale in halo formation that has wide-ranging applications across cosmology.
Naively, computing it requires 
repeated numerical integration to calculate the variance
of the power spectrum on different scales and determine which scales exceed the threshold for nonlinear collapse.
We accelerate this calculation by working 
in configuration space and approximating the correlation function as a polynomial at $r \leq 5$ $h^{-1}$ Mpc. This enables an analytic
rather than numerical solution,  
accurate across a variety of cosmologies
 to $0.1$--$1\%$ (depending on redshift) and $10$--$20\times$ faster than the naive numerical method.
We also present a further acceleration ($40$--$80\times$ faster than the naive method) in which we determine
the polynomial coefficients using 
a Taylor expansion in the cosmological parameters
rather than re-fitting a polynomial to the correlation function.
Our acceleration greatly reduces the cost of repeated calculation of the nonlinear mass. This will be useful for MCMC analyses to constrain cosmological parameters from the highly nonlinear regime, e.g.\ with data from upcoming surveys.
We make our \textsf{python} code publicly available at \url{https://github.com/akrolewski/NonlinearMassFaster}.

\end{abstract}



\section{Introduction}




Computing the spatial scale on which the density fluctuations have variance of order unity is a common problem in cosmology. In bottom-up structure formation, fluctuations are small on large scales and become progressively larger on smaller scales. 
As the density fluctuations approach unity, Perturbation Theory (PT)-plus-biasing-based models of the clustering \citep{Bernardeau:2002} break down,
and density fluctuations begin to collapse into dark matter halos.

The nonlinear scale $\Rnl$ is the characteristic
scale at which these processes occur.  Its technical definition is the scale at which the rms of the density field
fluctuations, $\sigma_R$, reaches $\dc = 1.686$ \citep{BryanNorman98,Child:2018}, the 
linear-density threshold for spherical tophat collapse \citep{GunnGott72}.
The nonlinear scale $\Rnl$ can also be converted into a nonlinear mass $\Mnl$ by multiplying by the background density $\rhobgd$.\footnote{The background density may be either the matter density or the critical density, but
 does not contain  a factor of $\Delta_{\rm c} \sim 200$, the final-state overdensity of a virialized halo. This is because the nonlinear mass is defined with reference to the {\it linear}
density threshold for the intial conditions of collapse.}

The nonlinear mass depends
weakly on cosmology,
with the cosmology dependence arising
from the small-scale power spectrum.
Beyond the trivial dependence
on the amplitude
$\sigma_8$ and the spectral slope $n_{\rm s}$,
$\Rnl$ is most
sensitive to $\Omega_{\rm m}$.
Increasing $\Omega_{\rm m}$ decreases the elapsed
time in radiation domination. This leads to less suppression
of small-scale modes entering the horizon during
radiation domination, ultimately increasing small-scale
power. 
The small-scale power spectrum is also sensitive
to $\Omega_{\rm b}$, both because baryons slow the growth
of structure after matter-radiation equality but
before decoupling \citep[equation E-6 in][]{Hu:1996},
and because the baryonic Jeans scale suppresses power at $k \geq 300$ $h$ Mpc$^{-1}$.


The nonlinear mass has broad applications across cosmology.
  Most importantly, it is the characteristic scale of halo
formation in a scale-free power-law cosmology \citep{Kravtsov12} and is consequently
the characteristic mass scale for self-similar scaling relations in galaxy clusters
\citep{Kaiser86,BryanNorman98,Norman10,Kravtsov12}.
Although the nonlinear mass is not exactly the characteristic halo mass scale in a  $\Lambda$CDM power spectrum,
it is a good enough approximation that deviations from self-similar scalings are often
parameterized in terms of it \citep{Kravtsov12}. As an important determinant of halo formation,
the nonlinear mass is the key scale in the halo growth rate \citep{Wechsler02}, the
concentration-mass relation \citep{Child:2018}, assembly bias \citep{Dalal08}, the mass-bias relation \citep{SeljakWarren04},
and spin alignment between halos and filaments \citep{Hahn07a,Hahn07b}.  

Due to these broad applications, the nonlinear
mass plays a role in modeling nonlinear structure growth \citep{Abazajian05} and parameterizing the halo mass function
to obtain $\sigma_8$ \citep{Seljak05}.  Prompt calculation of the nonlinear mass will allow its inclusion in MCMC chains, enhancing cosmological constraints from the highly nonlinear regime, including weak lensing halo mass profiles \citep{Umetsu19} and cluster abundances \citep{Bocquet19}.  
For instance, in the baryon-feedback model of \citet{Mead15}, which is included in the KiDS
weak lensing analysis \citep{Hildebrandt18},
the amplitude of feedback depends on the mass-concentration relation,
which is in turn dependent on the nonlinear mass \citep{Child:2018}.



We now outline how one might naively calculate the nonlinear mass, explain why this is inefficient, and sketch the approach of this work to accelerating the calculation. 
In the naive method, one computes a numerical
integral for $\sigma_R$ at each point in $R$
traversed by a numerical root-finder solving the equation  $\sigma_R = \dc$.
The combination of the numerical integration and the root-finding
makes $\Rnl$ slow to calculate.  

In this work we present a scheme to greatly accelerate this calculation.
Our method uses the algebraic solution of a cubic equation to determine $\Rnl$, thereby bypassing
 both the numerical
integration and the root-finding.
In particular, we work in configuration
space and fit a polynomial to the correlation function
on small scales.
 $\sigma_R$  is a compactly-supported integral over the correlation function, so these fitted coefficients immediately give us the integral's value as a cubic in $\Rnl$. The equation $\sigma_R = \dc$ can then be solved analytically.
 Our
method is accurate to $<1\%$ for a variety of cosmologies, an order of magnitude faster than the standard method, and can be further accelerated by
an additional 4$\times$ using a Taylor series to determine the polynomial coefficients
of the correlation function.





All numerical work in this paper uses the best-fit cosmology from the {\it Planck} 2018 release \citep{Planck:2018}
with $\Omegam = 0.3096$, $\Omegab = 0.04897$, $\ns = 0.9665$, $\sigma_8 = 0.8102$,
and
$h = 0.6766$.\footnote{If one computes $\sigma_8$
using our fiducial CDM-plus-baryons $P(k)$, one obtains
0.8138, in contrast to the {\it Planck} value of 0.8102,
which is calculated for $P(k)$ including CDM, baryons
and neutrinos.}
Consistent with past work \citep[e.g.][]{Child:2018},
we use the linear power spectrum
of cold dark matter plus baryons, since halos do not respond
to neutrinos \citep{Costanzi13,VN14,Castorina14,Castorina15}.

\section{Method and implementation}

In this section,
we review the calculation of the variance
of the density field in \S\ref{sec:variance},
present our algebraic method in \S\ref{sec:solve_Rnl},
and show the solution to the cubic in \S\ref{sec:details}.

\subsection{Variance of the density field}
\label{sec:variance}

The variance of the linear density field at a point $\vec{x}$ and  redshift $z$
 within a sphere of radius $R$ is
 \begin{align}
&\sigma^2_R(\vec{x},z) = 
\label{eqn:sigmasq_def}
\\
&V_R^{-2}\int d^3\vec{r}\;
d^3\vec{r}'\;\Theta(R-|\vec{r}|)
\Theta(R-|\vec{r}'|)\delta_{\rm lin}(\vec{x}+\vec{r},z)\delta_{\rm lin}(\vec{x}+\vec{r}',z).\nonumber
\end{align}
$\Theta$ is a Heaviside function, unity where its argument is positive and zero otherwise. 
In 3-D, the Heaviside function of radius
is simply a spherical tophat.
$V_R = 4 \pi R^3/3$ is the volume of a sphere of radius $R$.

The statistical homogeneity of the density field implies
translation invariance, and we may therefore write the
average over $\vec{x}$ as
\begin{align}
\sigma_R^2(z) &\equiv \left<\sigma_R^2(\vec{x},z) \right>  =
\frac{1}{V}\int d^3\vec{x}\;\sigma_R^2(\vec{x},z)\nonumber\\
& = V_R^{-2}\int d^3\vec{r}\;d^3\vec{s}\;\Theta(R-|\vec{r}|)
\Theta(R-|\vec{r} + \vec{s}|)\xi(s,z) 
\label{eqn:sigmasq_homogeneity}
\end{align}
where $\vec{s}=\vec{r}'- \vec{r}$,
with $\xi$ is the {\it linear} matter correlation function:
\begin{equation}
    \xi(s,z) \equiv \int d^3\vec{x}\;\delta_{\rm lin}(\vec{x}+\vec{r},z);\delta_{\rm lin}(\vec{x}+\vec{r}+\vec{s},z)
\label{eqn:xi_def}
\end{equation}
and $\delta_{\rm lin}$ is the linear density field.
To obtain the second equality in equation~(\ref{eqn:sigmasq_homogeneity}), we inserted equation~(\ref{eqn:sigmasq_def}) for $\sigma_R(\vec{x},z)$
and integrated over $d^3\vec{x}$, first using the definition of $\xi$ in equation~(\ref{eqn:xi_def}).
By recasting equation~(\ref{eqn:sigmasq_homogeneity}) as a convolution we obtain
\begin{align}
\sigma_R^2(z) 
= V_R^{-2}\int d^3\vec{s}\;\xi(s,z)\left[\Theta(R)\star \Theta(R) \right](\vec{s})
\label{eqn:tophat_config}
\end{align}
where ``star'' denotes convolution.\footnote{This formula offers a geometric way to show that the overlap integral of two spherical Bessel functions $j_1(kR)j_1(ks)$ scales as the volume of the lens formed by the overlap of two spheres.}
The convolution inside the square brackets is evaluated at an offset $\vec{s}$ and is itself an integral over the dummy variable $\vec{r}$; for clarity we have suppressed this latter argument.
Equation~(\ref{eqn:tophat_config}) shows that the variance is thus just the integral of the correlation function against a kernel given by the convolution of two spherical tophats  at an offset $\vec{s}$.

The overlap of the two spheres
forms a lens.
Consequently we can evaluate the convolution using the formula for the volume of a lens  produced by overlapping two spheres of radius $R$, offset from each other by $s$ \citep{Weisstein:sphere}:
\begin{align}
\left[\Theta(R)\star \Theta(R) \right](\vec{s}) = \Vlens(s; R) = \frac{\pi}{12} (4R + s) (2R- s)^2. 
\label{eqn:lens}
\end{align}
    In the limit $s \rightarrow 0$, i.e.\ when the two spheres share a common center, this expression recovers
    the volume of a sphere.

\begin{figure}
    \centering
    \includegraphics[width=9cm]{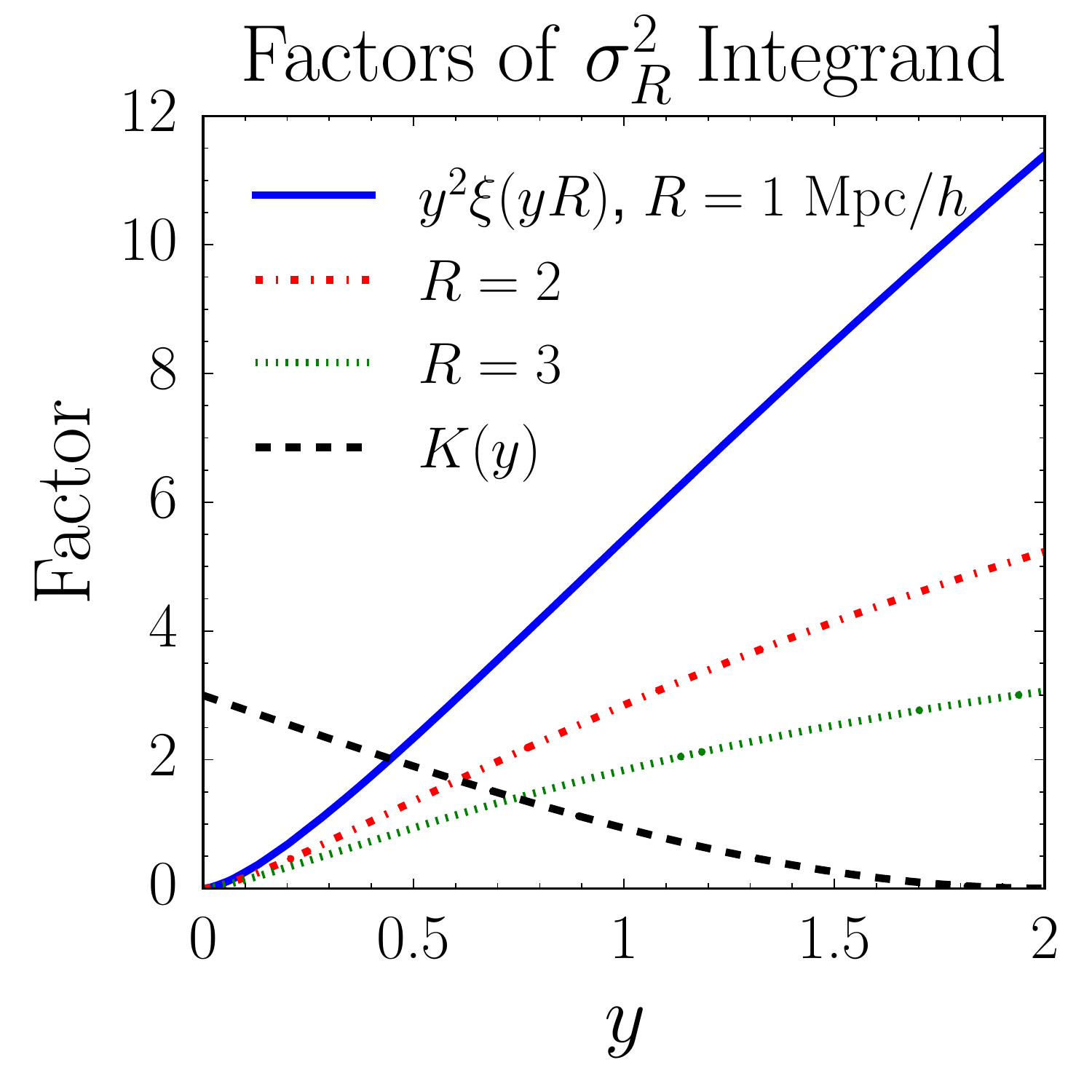}
    \caption{Different pieces of the integral (\ref{eqn:dropped}). The dashed black line is the kernel $K(y)$ defined in equation~(\ref{eqn:kernel}); and the short-dashed, dot-dashed and solid lines (various colors) are $y^2\xi(yR)$ at $R = 1,\;2,$ and $3\;\Mpch$.
    The dominant contribution to $\sigma^2_R$
    comes from intermediate $y$, where both $K(y)$ and $y^2 \xi(yR)$ are nonzero.
   }
    \label{fig:xiy_kernel}
\end{figure}

Inserting equation~(\ref{eqn:lens}) in equation~(\ref{eqn:tophat_config}) yields
\begin{align}
\sigma_R^2(z) &= 4\pi V_R^{-2}  \int_0^{2R} s^2 ds\; \xi(s,z)\Vlens(s;R)   \nonumber\\ 
&= 4\pi V_R^{-2} \int_0^{2R} s^2 ds\; \xi(s,z) (4R + s)(2R - s)^2 \nonumber\\ 
&=\frac{\pi^2R^3}{3V_R^2} \int_0^2 y^2 dy\; \xi(yR,z) (4R + yR)(2R - yR)^2.  
\label{eqn:sigmasq_vlens}
\end{align}
To obtain the third equality we changed variables to $s = yR$, $s^2 ds = R^3 y^2 dy $. Simplifying the last line 
we obtain the formula of \citet{Zehavi:2005}, quoted
there without proof but obtained by direct integration:\footnote{D. Eisenstein, personal communication.}
\begin{align}
\sigma_R^2(z) =\int_0^2 dy\;y^2\;\xi(yR,z)\; K(y), 
\label{eqn:dropped}
\end{align}
with
\begin{equation}
K(y) = \left(3-\frac{9y}{4} + \frac{3y^3}{16} \right). 
\label{eqn:kernel}
\end{equation}
We will refer to $K(y)$ as defined in equation~(\ref{eqn:kernel}) as the ``kernel'' for the remainder of this work.  We plot the components of equation~(\ref{eqn:dropped}) in Figure~\ref{fig:xiy_kernel}, including
the kernel and $y^2 \xi (y R,z=0)$ for different values of $R$.
Since $K(y)$ is nearly zero at $y > 1.5$ and $y^2 \xi(yR)$
is nearly zero at $y < 0.5$, much of the integral comes from
intermediate values of $y$.

\subsection{Solving for the nonlinear scale}
\label{sec:solve_Rnl}

 The standard approach to computing $\Rnl$ is to use the Convolution Theorem to perform the convolution (equation~\ref{eqn:tophat_config}) as a product in Fourier space, i.e.
\begin{align}
\sigma_R^2(z) = \int \frac{k^2 dk}{2\pi^2} \left[\frac{3j_1(kR)}{kR}\right]^2 P(k,z),
\label{eqn:fourier_space}
\end{align}
where the quantity in square brackets is the square of the Fourier Transform
of a spherical tophat.
One would then
use numerical root-finding to solve the equation $\sigma_{R_{\rm NL}}(z) = \dc = 1.686$.

  Our method evaluates $\sigma^2_R$ from equation~(\ref{eqn:dropped}) in configuration space
  and fits a low-order polynomial to the small-scale
  correlation function, leading to an analytic
  integral that enables algebraic calculation of $\Rnl$. This is an order of magnitude
  faster than the standard method to compute $\Rnl$
  from numerical integrals of the power spectrum.
  
\begin{figure}
    \centering
    \includegraphics[width=0.5\textwidth]{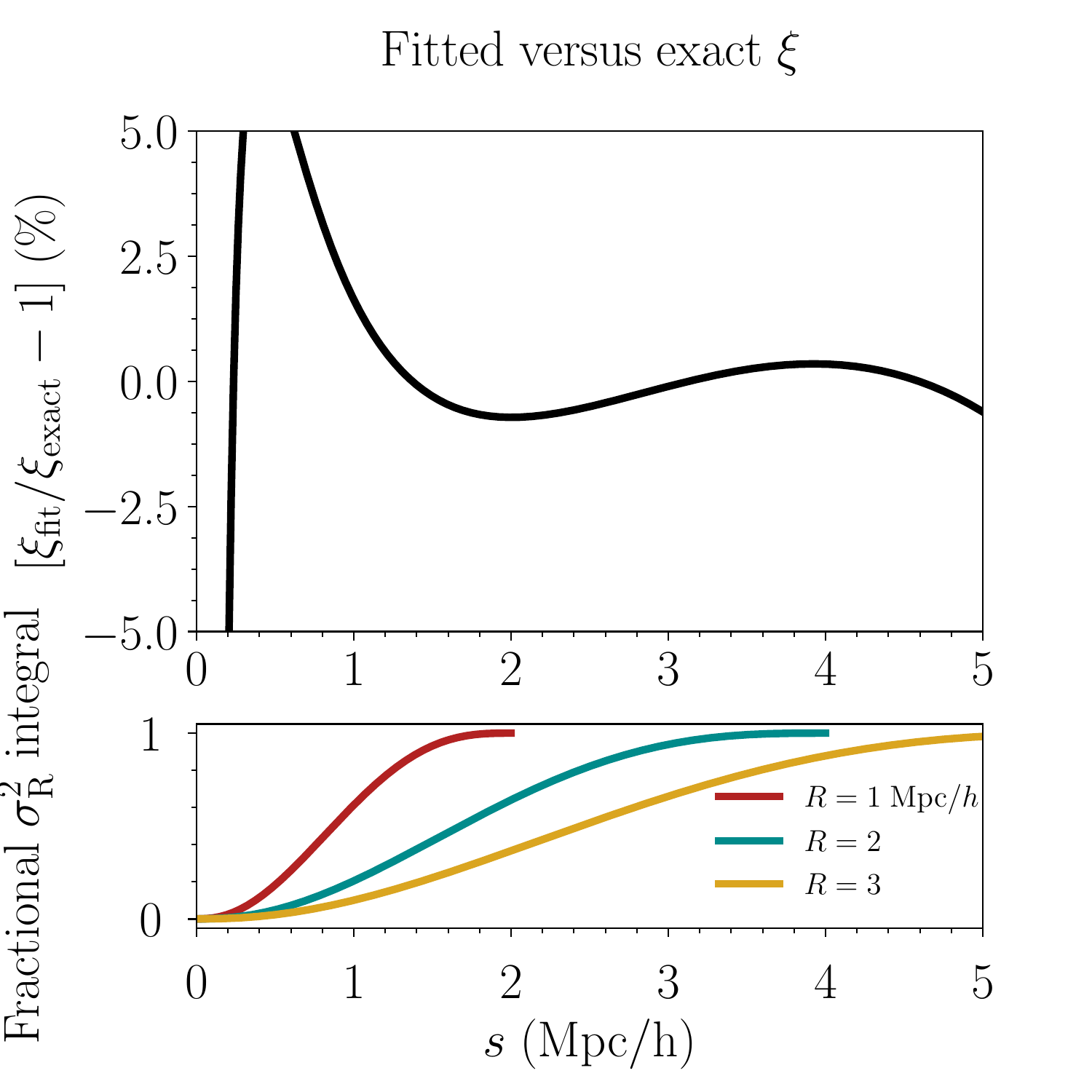}
    \caption{  \textit{Top---} Difference between $s^2\xi_{\rm lin}(z=0)$
    and the third-order polynomial fit to it. The differences in fitting quality
    are negligible among different cosmologies.
    We recover the exact correlation function to rather high accuracy, generally sub-percent over most of its domain; larger errors occur at small scales which contribute little to the fractional $\sigma^2_R$ integral, as shown in the bottom panel.
    \textit{Bottom---} Fractional
    buildup of $\sigma^2_R$ (equation \ref{eqn:dropped})
    as a function of $s \equiv yR$. Since the integral extends to $y=2$, the curves
    cut off at respectively $yR = 2,\; 4$ and $6$ 
    for $R = 1,2$, and $3\;\Mpch$.  For a wide range in $R$,
    the integral turns out to be most sensitive to exactly the region in which our fit performs best, $1 < s < 5$ $h^{-1}$ Mpc.
    }
    \label{fig:xi_fit}
\end{figure}

  Our method offers two advantages that substantially accelerate the calculation of the nonlinear scale.
First, the configuration space integral is easier to handle
than the Fourier space integral,
which has an infinite upper bound
and BAO wiggles which require a larger number of $k$
steps for accurate sampling.
Second, the correlation function on small scales is smooth
and can be approximated
by a low-order polynomial (Figure~\ref{fig:xi_fit}).
This allows the $\sigma^2_R$ integral to be evaluated
 analytically and $\Rnl$ then computed algebraically.

To obtain an analytic expression for $\sigma^2_R$
for a polynomial correlation function,
we start with the center line in equation~(\ref{eqn:sigmasq_vlens}); separating the integrals term by term we find
\begin{align}
\sigma_R^2(z) = D^2(z)\;R^{-3}\bigg\{&3\int_0^{2R}ds\;s^2 \xi(s) -\frac{9}{4R}\int_0^{2R}ds\;s^3\xi(s) \nonumber\\
&+ \frac{3}{16R^3}\int_0^{2R} ds\;s^5\xi(s)\bigg\},
\label{eqn:three_terms}
\end{align}
where we explicitly separate the redshift-dependent piece of the linear correlation function,
the square of the linear growth factor
$D^2(z)$,
and hereafter use $\xi(s)$ to mean $\xi(s,z=0)$.

We perform the integral in equation~(\ref{eqn:three_terms}) analytically
by expanding $s^2 \xi(s)$ as a polynomial:
\begin{align}
s^2 \xi(s) = \sum_{n=0}^{n_{\rm max}}c_n s^n. 
\label{eqn:poly}
\end{align}
We chose a polynomial because it allows the integral
in equation~(\ref{eqn:dropped}) to be done analytically
and provides a good fit to the correlation function
over the restricted range required
($s \leq 5$ $h^{-1}$ Mpc).

Inserting the expansion~(\ref{eqn:poly})
into equation~(\ref{eqn:three_terms}) and performing the integrals,
we obtain the following expression for $\sigma^2_R$,
which we set equal to $\dc^2$:
\begin{align}
\sigma_{\Rnl}^2(z) &= D^2(z)\sum_{n=0}^{n_{\rm max}} c_n \Rnl^{-3}\bigg\{3 \; \frac{2^{n+1}}{n+1}\Rnl^{n+1} - \frac{9}{4\Rnl}\frac{2^{n+2}}{n+2} \Rnl^{n+2}\nonumber\\
&\qquad \qquad \qquad \qquad \; \; +\frac{3}{16\Rnl^3}\frac{2^{n+4}}{n+4} \Rnl^{n+4} \bigg\}  
\nonumber\\ 
&= D^2(z)\sum_{n=0}^{n_{\rm max}}  2^{n+1} c_n R_{\textrm{NL}}^{n-2} \bigg\{\frac{9}{n^3 + 7n^2 + 14n + 8}\bigg\} = \dc^2.
\label{eqn:poly_sum}
\end{align}
We transform equation~(\ref{eqn:poly_sum}) from a sum
of inverse powers to a polynomial
by multiplying through by $\Rnl^2$:
 \begin{align}
    \frac{\Rnl^2 \dc^2 }{D^2(z)} =  \sum_{n=0}^{n_{\rm max}}  2^{n+1} c_n \Rnl^{n} \bigg\{\frac{9}{n^3 + 7n^2 + 14n + 8}\bigg\} 
    \label{eqn:sigmasq_poly}
 \end{align}
Quartics and lower-order polynomials have
 a closed form solution, but quintics and higher-order
 polynomials do not (this is known as the Abel-Ruffini theorem).
 Therefore,
if $n_{\rm max}\leq 4$, we can solve in closed form for $\Rnl$.

\begin{figure}
    \centering
    \includegraphics[width=0.5\textwidth]{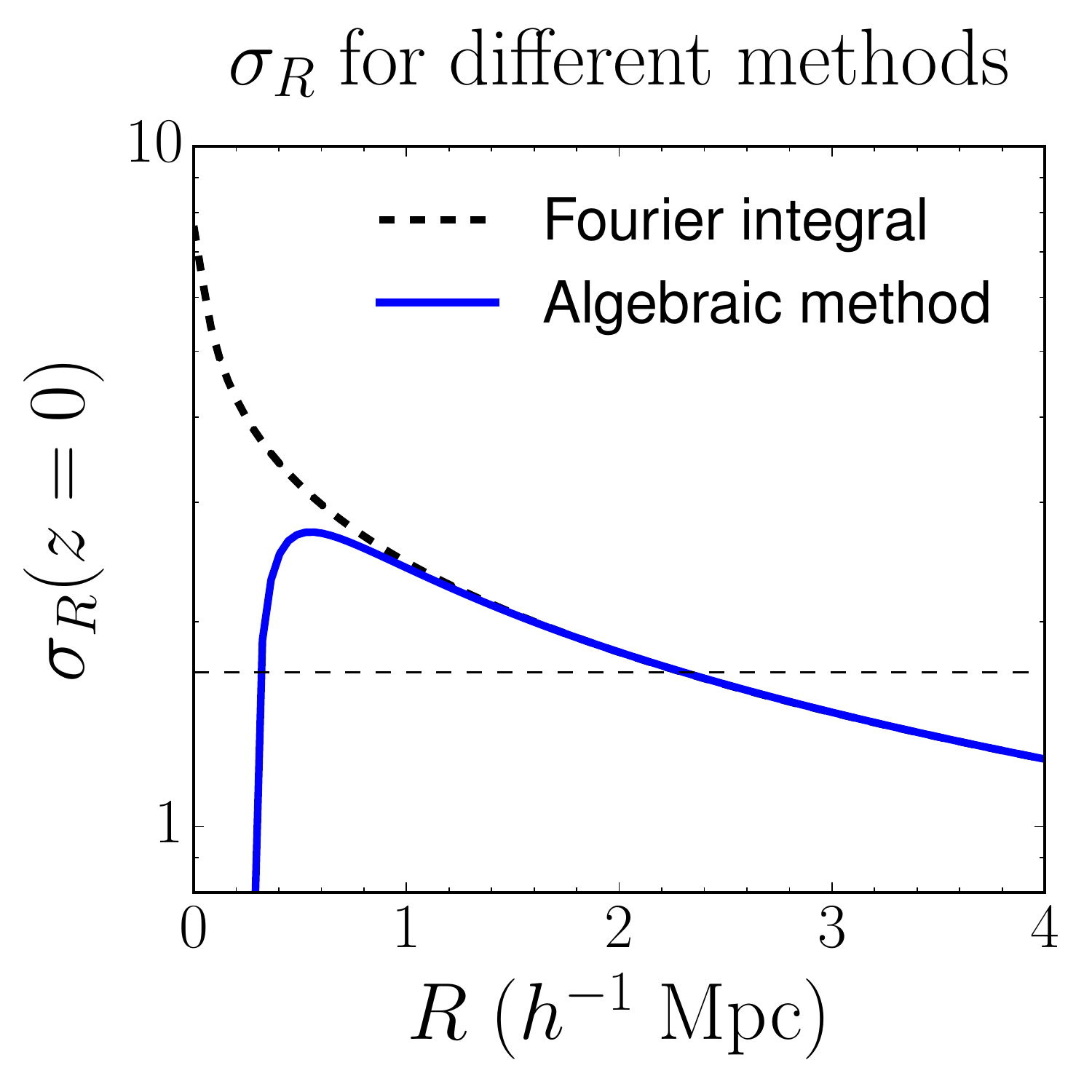}
    \caption{Comparison between exact $\sigma_R$ from the Fourier integral (blue; equation~\ref{eqn:fourier_space}) and our algebraic method (black; equation~\ref{eqn:cubic}).
    Both calculations are done at $z=0$; results for $\sigma_R(z)$ will be very similar except that the horizontal axis will be scaled by $R_{\rm NL,fid}(z)/R_{\rm NL,fid}(z=0)$.  The thin black dashed line gives
    $\sigma_R = \dc = 1.686$; thus $\Rnl$ is where the thicker curves cross the dashed line. While the solid and dashed curves disagree at small scales, the agreement is almost perfect near where $\sigma_R = 1.686$.
    }
    \label{fig:sigmaR}
\end{figure}

We find that $n_{\rm max}=3$ is sufficient 
to reproduce $s^2 \xi(s)$ to percent-level
accuracy 
(Figure~\ref{fig:xi_fit}), implying
that an algebraic solution for $\Rnl$ exists.
While Figure~\ref{fig:xi_fit} only shows the fit to the correlation function in the {\it Planck} 2018 cosmology, $s^2 \xi(s)$ can be approximated
equally well by a cubic across a
wide range of cosmologies.
Because the cubic provides a good fit across
a wide range in $s$, the analytic
approximation of $\sigma_R$
from integrating equation~(\ref{eqn:three_terms})
provides a very good match
to the numerical solution 
from the Fourier-space integral at $R > 0.5$ $h^{-1}$ Mpc at $z=0$ (Figure~\ref{fig:sigmaR}).

We found that the cubic provides the best balance
between simplicity and accuracy: a quadratic approximation
is considerably less accurate, whereas a quartic
offers only minimal improvement.
Other possibilities, such as omitting the constant and linear
terms or requiring the constant term to be positive,
degrade the accuracy of the fit.
While a piecewise function (e.g.~a smoothing spline) can reproduce $s^2 \xi(s)$
to arbitrary accuracy, the upper bounds in the integrals in equation~(\ref{eqn:three_terms})
are no longer linear multiples of $R$, and thus contribute an $R^{-6}$ term
in equation~(\ref{eqn:poly_sum}). If $n_{\rm max} \geq 1$,
this yields a 5th-order polynomial with no analytic solution
in equation~(\ref{eqn:cubic}).

Equation~(\ref{eqn:poly}) must approximate $s^2 \xi(s)$
well at $s < 2\Rnl(z)$, since this is the upper bound of the 
integral in equation~(\ref{eqn:three_terms}).
To avoid the circularity of requiring $\Rnl$ to fit the $c_n$,
we fit the $c_n$ to $s < 1.9 R_{\rm NL,fid}(z)$,
where $R_{\rm NL,fid}$ is $\Rnl$ in the fiducial
{\it Planck} 2018 cosmology. We empirically
find that using an upper cutoff of $1.9 R_{\rm NL,fid}(z)$ leads to 20\% better accuracy
than using $2 R_{\rm NL,fid}(z)$. This slightly
up-weights smaller and intermediate
scales which contribute more to the $\sigma^2_R$ integral (bottom panel of Figure~\ref{fig:xi_fit}).

Allowing the fitting range (and thus the $c_n$) to vary in redshift is critical,
because the error on the cubic increases greatly at small $s$:
this is 
the sharp drop in the signed deviation
between the correlation function and the cubic fit at $s \leq 0.5$ $h^{-1}$ Mpc in Figure~\ref{fig:xi_fit} (or equivalently,
the downturn in $\sigma^2_R$ at $R > 1$ $h^{-1}$ Mpc in Figure~\ref{fig:sigmaR}).
If we calculated $\Rnl(z=6)$ using $c_n(z=0)$, we 
would be primarily using scales where the cubic provides
an extremely poor fit to $s^2 \xi(s)$, leading to a severe
loss of accuracy. Instead, we fit over a very restricted
$s$ range at $z=6$, ensuring an accurate fit over the vast majority of the relevant range in $s$.
Because of this rescaling in $R_{\rm NL,fid}(z)$, 
Figures~\ref{fig:xi_fit} and~\ref{fig:sigmaR} at $z=0$
are very similar to the equivalent figures at high
redshift, except that the $s$ ($R$) axis
will be rescaled by $R_{\rm NL,fid}(z)/R_{\rm NL,fid}(z=0)$.

\subsection{Detailed solution of the cubic}
\label{sec:details}



In this section we explicitly show how one obtains $\Rnl$ algebraically.
Evaluating equation~(\ref{eqn:sigmasq_poly}) with $n_{\rm max} = 3$ yields
\begin{align}
\frac{9}{4} R^{-2} c_0 + \frac{6}{5} R^{-1} c_1 + c_2 + \frac{36}{35} R \, c_3 = \frac{\dc^2}{D^2(z)}.
\label{eqn:cubic}
\end{align}
To simplify what follows, we define coefficients $a_i$ that incorporate both the $c_n$ and their numerical pre-factors in equation~(\ref{eqn:cubic}), as
\begin{align}
a_0 = \frac{9c_0}{4},
\quad
a_1 = \frac{6c_1}{5},
\quad
a_2 = c_2,
\quad
a_3 = \frac{36c_3}{35}.
\end{align}
We now rewrite equation~(\ref{eqn:cubic}) in standard cubic form
\begin{align}
R^3 + \alpha_2 R^2 + \alpha_1 R + \alpha_0 = 0
\end{align}
with the $\alpha_i$ given as
\begin{align}
\alpha_0 = \frac{a_0}{a_3},
\quad
\alpha_1 = \frac{a_1}{a_3},
\quad
\alpha_2 = \frac{a_2-1/D(z)^2}{a_3}.
\end{align}
We can obtain the solution using Cardano's formula for the cubic \citep{Weisstein:cubic}. We find the roots $R_i$ as:
\begin{align}
    R_1 = - \frac{1}{3} \alpha_2 + (S+T), \nonumber
\end{align}
\begin{align}
    R_2 = - \frac{1}{3} \alpha_2 - \frac{1}{2} (S + T) + \frac{1}{2} i \sqrt{3} (S - T), \nonumber
\end{align}
\begin{align}
        R_3 = - \frac{1}{3} \alpha_2 - \frac{1}{2} (S + T) - \frac{1}{2} i \sqrt{3} (S - T).
\end{align}
We define the auxiliary variables $Q$, $R$, $D$, $S$, and $T$ as
\begin{align}
    Q \equiv \frac{3 \alpha_1 - \alpha_2^2}{9},
    \quad
    R \equiv \frac{9 \alpha_2 \alpha_1 - 27 \alpha_0 - 2 \alpha_2^3}{54}, \nonumber
\end{align}
\begin{align}
    D \equiv Q^3 + R^2,
    \quad
    S \equiv \sqrt[3]{R + \sqrt{D}},
    \quad
    T \equiv \sqrt[3]{R - \sqrt{D}}.
\end{align}
We choose the real and positive root.
Note that despite the presence
 of $i$ in $R_2$ and $R_3$, they need not be complex because $S$ and $T$ are also complex and can render the factor involving $i$ real overall.

\section{Numerical implementation}
\label{sec:implementation}

To obtain the nonlinear scale using the method outlined in \S\ref{sec:solve_Rnl} and \S\ref{sec:details}, we first need the linear correlation function. We obtain the correlation function by transforming the linear power spectrum from CAMB \citep{Lewis:1999bs,Howlett:2012mh}.\footnote{\url{http://camb.info}} We use 800 logarithmically-spaced sample points per decade over the range $k = 10^{-3}$ to $10^4$ $h$ Mpc$^{-1}$. To transform from $P(k)$ to $\xi(s)$ we use
 $2 \times 10^4$ $s$-points evenly spaced between $s = 0$ and 5 $h^{-1}$ Mpc.
For consistency with past work \citep{Child:2018}, we use the power spectrum of baryons plus CDM in our numerical implementation,
but the method is general and can accept 
an arbitrary linear power spectrum as input.

For the timing tests in \S\ref{sec:timing_and_accuracy},
we find $\Rnl$ using optimized implementations of both the algebraic method
and the numerical integral methods in Fourier and configuration space.
We aim to achieve $0.01\%$ accuracy on $\Rnl$
for the numerical integral methods
in both Fourier and configuration space.


The most time-consuming part of our implementation of the polynomial method
is solving for the polynomial coefficients $c_n$.  We use 1000 sampling points for the 
correlation function and determine $c_n$ via linear least-squares.
The relevant operations are vectorized so the performance is not highly sensitive to the number of sampling points. Consequently we can choose 1000 points to preserve accuracy yet not pay much price in speed.
The rate-limiting step is matrix inversion in the linear least squares fitting; to speed this up, we take advantage of the fact that the relevant matrix is symmetric,
as it is the product of the Vandermonde matrix and its transpose.
We solve the least squares equation using the \textsf{lapack} linear algebra package
routine \textsf{dpotrs} ({\bf d}ouble-precision  {\bf po}sitive {\bf tr}iangular matrix {\bf s}olve), which uses Cholesky decomposition to efficiently invert a symmetric matrix \citep{NumericalRecipes}.\footnote{\url{http://www.netlib.org/lapack/}}
This approach is considerably faster than the \textsf{numpy} least-squares package, both by eliminating \textsf{numpy}
overheads and by using a faster method specifically
appropriate for symmetric matrices.

As an alternative to the time-consuming determination of the polynomial coefficients,
we determine $c_n$ using a Taylor expansion in
the cosmological parameters ($\Omega_{\rm m}$, $\Omega_{\rm b}$, $n_{\rm s}$) centered on the fiducial cosmology.
The dependence on $\sigma_8$ is trivial, as it just
rescales the polynomial coefficients by a multiplicative
factor.
This expansion assumes the mapping of parameters
to power spectra appropriate for a CDM cosmology;
modifications to the transfer function,
e.g. by warm dark matter or oscillations in the inflationary potential,
will change this mapping
and require fitting the polynomial coefficients rather
than using the Taylor series.

To compute the Taylor expansion for each $c_n$, we first compute
correlation functions and $c_n$ for four cosmologies per parameter varied ($\Omega_{\rm m}$, $\Omega_{\rm b}$ and $n_{\rm s}$): two with the parameter varied by $\pm$1$\sigma$
from the {\it Planck} 2018 best-fit value (with $\sigma = 0.0056$, $0.001$ and $0.0038$, respectively) and two with the parameter varied by $\pm$5$\sigma$.
Then for each $c_n$ and parameter $p$, we fit a line $c_n(p)$
with the intercept fixed to reproduce $c_n$ in the fiducial cosmology.
This allows us to achieve a good fit for a broad range of cosmologies
away from the {\it Planck} best-fit cosmology.

We must re-fit $c_n$ at each $z$
to ensure an accurate fit over most of the relevant
range in scale (\S\ref{sec:solve_Rnl}).
Therefore,
we must also determine the Taylor coefficients
as a function of redshift.
We measure the first-order
Taylor coefficients for the
three parameters for 60 sampling redshifts
spaced at $\Delta z = 0.1$
between $z = 0$ and 6.
We determine the Taylor coefficients
at arbitrary $z$ using a step
function taking each $z$ to the nearest
$\Delta z = 0.1$ grid point less than $z$.

Once the polynomial
coefficients are fit,
finding the nonlinear scale
is straightforward, requiring
only algebraic operations.
Nevertheless, we make a number of optimizations to the code implementing this.
We store intermediate
calculations to reduce
computational expense,
use only built-in \textsf{python}
functions or functions
from the \textsf{math} library,
and use decimals
rather than fractions
wherever possible
to avoid an additional division.
We also optimize
the performance
 of the naive method,
 with numerical integration
 and root-finding (in either
 Fourier or configuration space),
 in order to provide a fair
 comparison.
 
 For the configuration
 space integral (equation~\ref{eqn:dropped}),
 we use 50 sampling points in $y$
and pre-compute the kernel
since it does not change from iteration to iteration of the root-finding.
We evaluate the integral
using direct summation over the 50 points in $y$. We find that this
gives sufficient accuracy (better than $10^{-4}$)
and is considerably faster
than second-order methods
such as Romberg integration.
As with fitting the polynomial
coefficients, the scaling
of the integral is relatively insensitive
to the number of sampling points
due to the vectorization
of most operations. Therefore, our
results will not change much if the number of sampling points changes.

We use the \textsf{scipy} implementation of Brent's method
(\textsf{brentq}) \citep{Brent73} to perform the root-finding,
with the initial range between 0 and 5 $h^{-1}$
Mpc, comfortably bracketing $\Rnl$ in all plausible
cosmologies.  We use this method because
we find it to be the most robust for root-finding.\footnote{Newton-Raphson root-finding is 30--50\%
faster than Brent's method, but less robust as the root-finding has no bounds.
Therefore, for higher $z$ where $\Rnl$ becomes small, the root finder may attempt
to evaluate $\sigma_R$ at negative $R$, which is of course unphysical.}

We follow largely
the same procedure
to evaluate the integral
in Fourier space (equation~\ref{eqn:fourier_space}),
except that here, we find
that first-order summation
is inaccurate and instead
use the second-order trapezoid
rule.
We also downsample
the original $k$-space grid
to achieve better performance,
finding that 180 logarithmically-spaced points between $k_{\rm min} = 10^{-3}$ $h$ Mpc$^{-1}$
 and $k_{\rm max} = 10^{4}$ $h$ Mpc$^{-1}$ are adequate.
 Again,
 the scaling with number of points
 is weak, so our timings do not
 depend strongly on the number of sampling points chosen.
 
 Finally, we make our \textsf{python} code publicly available at \url{https://github.com/akrolewski/NonlinearMassFaster}.
 
 \begin{figure*}
    \centering
    \includegraphics[width=1.0\textwidth]{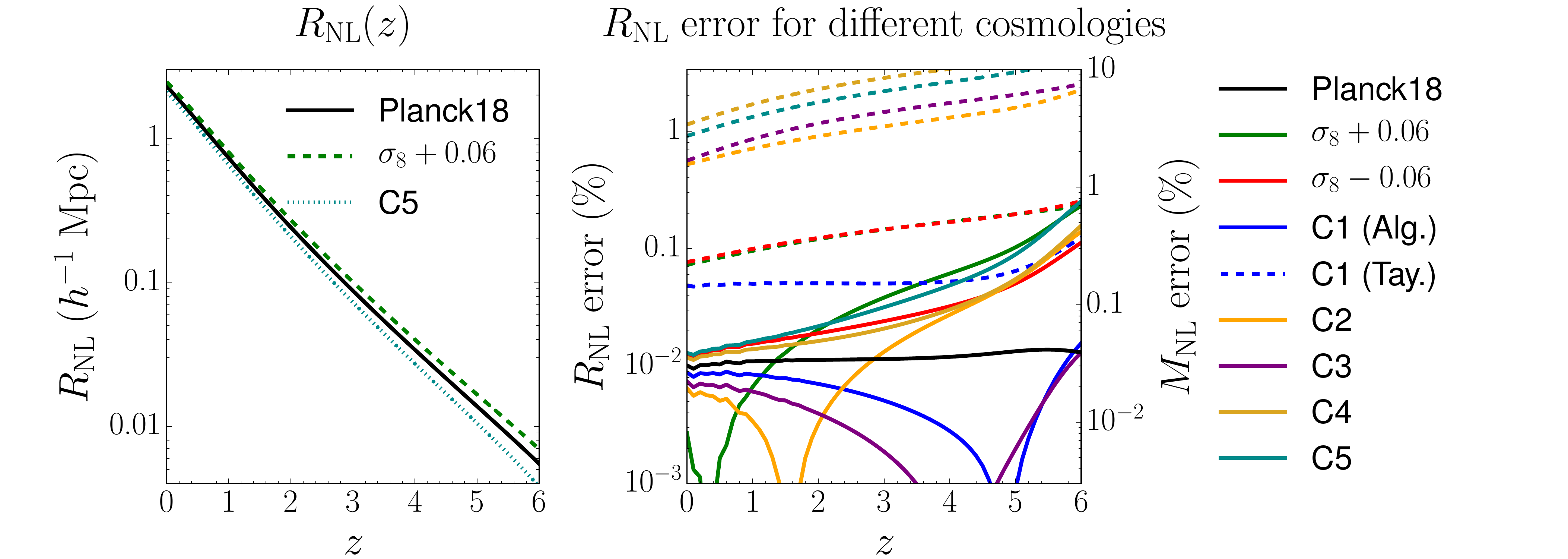}
    \caption{ {\textit Left:} $\Rnl(z)$ for the {\it Planck} 2018 cosmology (black), and the C5
    and $\sigma_8+0.06$ cosmologies (dark cyan, dotted; green, dashed), which have the largest discrepancy
    with the {\it Planck} 2018 $\Rnl$.
    {\it Right:} Absolute value of the error on $R_{\rm NL}$ (lefthand vertical axis) and the nonlinear mass (righthand vertical axis; 3$\times$ the $R_{\rm NL}$ error) as a function of redshift.
    We
    compare the accuracy for our 7 test cosmologies
    using the fiducial method, where we re-fit
    the coefficients to each cosmology (solid, colored curves). We also show the Taylor series
    method, where we use a Taylor expansion about the {\it Planck} 2018 cosmology to generate
    the coefficients (dashed curves).
    Each color stands for one of the 7 test cosmologies,
    and the line style indicates whether we re-fit
    the coefficients or determine them from the Taylor series.
    Cosmological parameters are given in Table~\ref{tab:cosmologies}.
   At $z \sim 1.5$, the error on $\Rnl$ for C2
changes sign from positive to negative, leading to the zero-crossing feature in the golden curve (and likewise
for the green, purple and blue curves at various redshifts).
    }
    \label{fig:accuracy}
\end{figure*}
 
\section{Results and discussion}
\label{sec:timing_and_accuracy}

We measure the accuracy
of our algorithm for finding $\Rnl$ compared
to numerical integration of equation~(\ref{eqn:fourier_space})
using a large number of sampling points in $k$.
We also measure the speed of our algorithm,
both when we fit the coefficients $c_n$ for every cosmology,
and when we use a Taylor expansion to calculate
$c_n$ for cosmologies sufficiently close to {\it Planck} 2018.
We compare the timing of our algorithm
to
optimized
versions of the numerical
integration and root-finding method
in both configuration and Fourier space.
We find the algebraic method is accurate to 0.1--1\% 
in mass
and offers a factor of 10--20 speedup
over the naive method, while
the Taylor series method is accurate to 1--10\% in mass
and 40--80$\times$ faster than the naive method.

We use \textsf{python} 3.5.2, with \textsf{numpy} 1.17.1 and \textsf{scipy} 1.3.1.
Timing tests are performed on a dual core 1.7 GHz Intel Core i5 processor.\footnote{This is a similar architecture to the Cori Haswell nodes at the National Energy Research Supercomputing Center (NERSC) (see \url{https://docs.nersc.gov/systems/cori/}),
hence the performance numbers outlined here can plausibly be scaled to get a rough estimate for the performance on a typical recent HPC system.} 
We use seven cosmologies to test our method. We start with two where
$\sigma_8$ is varied by $\pm 0.06$ from its best-fit {\it Planck} 2018 value, matching the tension between {\it Planck} and the low-redshift measurement from KiDS \citep{Hildebrandt18}.\footnote{Weak lensing
measurements are sensitive to the parameter
combination $S_8 = \sigma_8 \sqrt{\Omega_{\rm m}/0.3}$. KiDS measures $S_8 = 0.737^{+0.040}_{-0.036}$ \citep{Hildebrandt18};
if $\Omega_m$ is fixed to 0.3, this implies $\sigma_8 \sim 0.75$.}
We also use five test cosmologies
with $\Omega_{\rm m}$, $\Omega_{\rm b}$ and $n_{\rm s}$ drawn
from a random uniform distribution between 2$\sigma$ less than
and 2$\sigma$ greater than the {\it Planck} 2018 best-fit value
for each parameter. These explore a
a range in which many recent simulations lie
\citep{MultiDark,GLAM,Quijote}. The parameters for these test cosmologies
are given in Table~\ref{tab:cosmologies}.

  \begin{table}
    \centering
    \begin{tabular}{c|cccc}
Name & $\sigma_8$ & $\Omega_{\rm m}$ & $\Omega_{\rm b}$ & $n_{\rm s}$ \\
\hline
$\sigma_8$ $+$ 0.06 & 0.8702 & 0.3096 & 0.04897 & 0.9665 \\
$\sigma_8$ $-$ 0.06 & 0.7502 & --- & --- & --- \\
 {\it Planck} 2018 & 0.8102 & --- & --- & ---  \\

C1 & --- & 0.3129 & 0.0490 & 0.9669 \\
C2 & --- & 0.3185 & 0.0498 & 0.9676 \\
C3 & --- & 0.3145 & 0.0479 & 0.9616 \\
C4 & --- & 0.3086 & 0.0507 & 0.9591 \\
C5 & --- & 0.3004 & 0.0504 & 0.9663 \\
    \end{tabular}
    \caption{Cosmologies used to test our method.
    The default cosmology is {\it Planck} 2018,
    and the cosmologies C1--C5 are chosen
    by randomly drawing $\Omega_{\rm m}$, $\sigma_8$ and $n_{\rm s}$ from a uniform distribution of width $\pm 2 \sigma$ centered on the {\it Planck} 2018 values. Dashes indicate that a given parameter is unchanged from the row above.
    Accuracy and timing results for
    these cosmologies are given in Figures~\ref{fig:accuracy} and~\ref{fig:timing}.}
    \label{tab:cosmologies}
\end{table}

Our fiducial method, in which we re-fit the coefficients
for each input power spectrum, is accurate to better
than 0.3\% in $\Rnl$ (1\% in $M_{\rm NL}$)
at $0  < z < 6$
for all seven of these test cosmologies (Figure~\ref{fig:accuracy}).  If we instead
use a Taylor expansion about the {\it Planck} 2018 cosmology
to generate the coefficients, the
accuracy is somewhat worse, between 1\% and 10\% in mass.
Therefore, the Taylor series method
may be adequate at $z \sim 0$, where
it offers 1\% accuracy in $\Rnl$ and 3\% accuracy
in mass, but at higher redshifts it is likely
best to explicitly re-fit the polynomial coefficients,
depending on the accuracy demands of one's application.

We plot $R_{\rm NL}(z)$ in the left panel
of Figure~\ref{fig:accuracy} for the {\it Planck} 2018
cosmology ($R_{\rm NL,fid}$) and for the two
cosmologies with the largest deviation
in $R_{\rm NL}(z)$, C5 and $\sigma_8 + 0.06$.
At $z = 0$, $\Rnl$ in $\sigma_8 + 0.06$ (C5)
is 7\% higher (8\% lower) than $R_{\rm NL,fid}$,
increasing to 25\% higher (30\% lower) at $z=6$.
$\Rnl$ is very small at high redshift, with $\Rnl(z=6) \approx 0.005$ $h^{-1}$ Mpc. This does not imply
that the $z=6$ linear power spectrum is valid
out to $k_{\rm NL} = 2\pi/0.005 \approx 1000$ $h$ Mpc$^{-1}$.
Rather, the $k$ at which the linear and nonlinear
power spectra deviate is smaller than $k_{\rm NL}$
by a factor of a few.

Because we use $\Rnl$ in the {\it Planck} 2018
cosmology to set the fitting range,
the accuracy of our method
should degrade as the cosmology varies.
However, for certain
cosmologies and redshift ranges ($\sigma_8-0.06$
at $z < 1$, C2 at $z < 3$, and C1 and C3
at $z < 6$), the accuracy is {\it better}
than the accuracy for {\it Planck} 2018.
This is because the mismatch between
$\Rnl$ in these cosmologies and $R_{\rm NL,fid}$
is actually beneficial, since the (cosmology-dependent) optimal upper bound is not exactly $1.9 \Rnl$.
For these cosmologies and redshifts, $1.9 R_{\rm NL,fid}$ approaches the optimal upper bound at the points where the error approaches zero (e.g.\ $z\approx1.5$ for C2).

The $\Rnl$ error for many of the test cosmologies
increases at higher redshift. This is because
$\Rnl$ is smaller at high redshift, and 
thus
sensitive to the linear power spectrum at higher $k$
where the power spectrum is more cosmology dependent.\footnote{Changing $\Omega_{\rm m}$ and $n_{\rm s}$ tilts $P(k)$ with a pivot at $k \approx 0.1$ $h$ Mpc$^{-1}$. Therefore,
the linear power spectrum is more sensitive to cosmological parameters
at higher $k$.}
This means that the disagreement between $\Rnl$
and $R_{\rm NL,fid}$ is larger,
leading to a suboptimal fitting range and 
inaccurate polynomial coefficients.
As a consequence, if we instead solve for the scale
where $\sigma_R = 1$ \citep[e.g.][]{Norman10},
our method will be more accurate because this scale is larger and 
 less dependent on high $k$,
 yielding a better fitting range and more accurate
 polynomial coefficients.

\begin{figure}
    \centering
    \includegraphics[width=8cm]{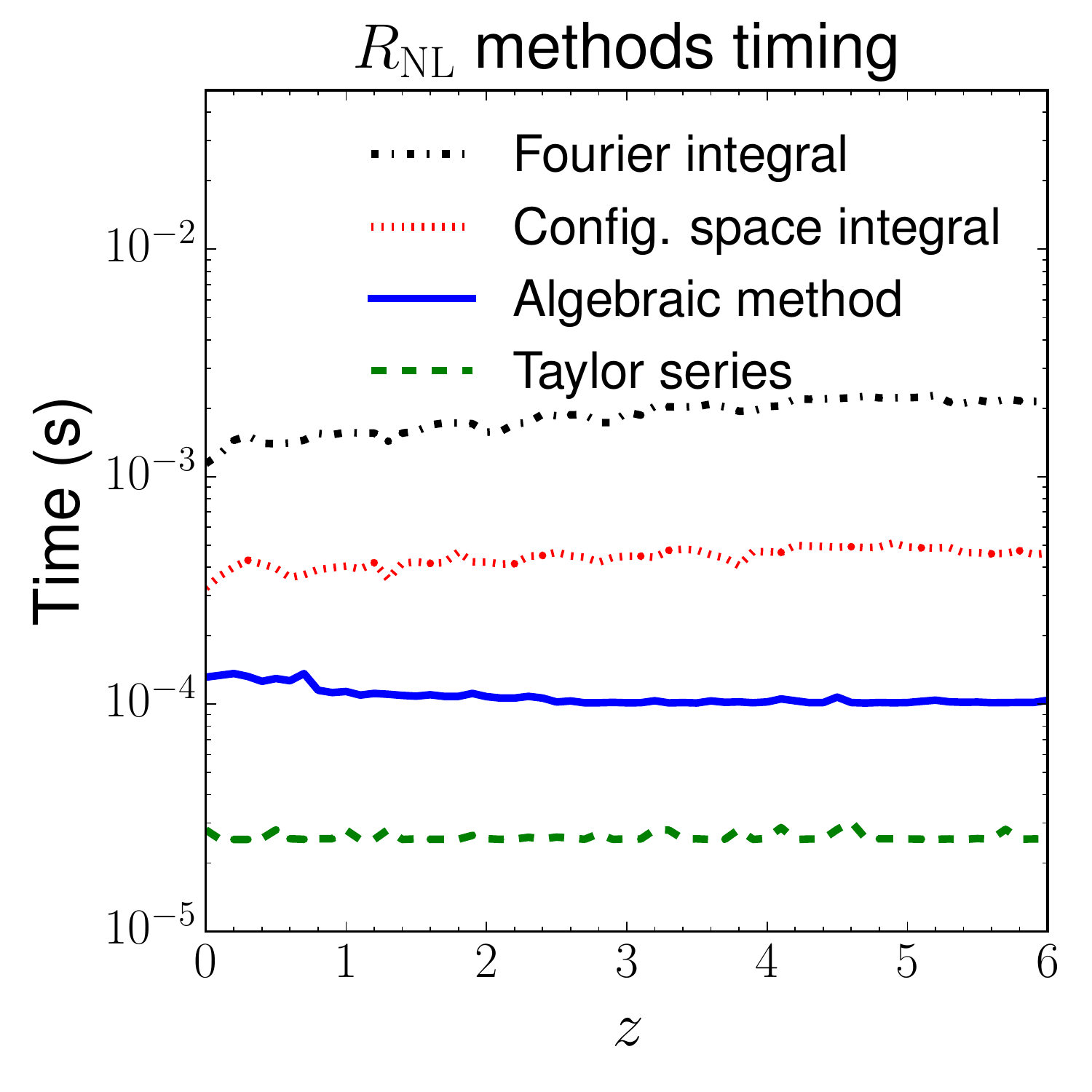}
    \caption{Comparison between
    the timing for our implementation
    of the algebraic method (blue)
    versus the numerical integration
    plus root-finding method either in Fourier
    space (black) or configuration space (red).
    We also show the timing of our method
    if we instead use a Taylor expansion
    to generate the coefficients
    rather than fit them to the correlation function (green).
    The algebraic method is 10--20$\times$ faster
    than the naive method with the integral
    in Fourier space, and the Taylor series method
    is 40--80$\times$ faster.
    }
    \label{fig:timing}
\end{figure}

We compare the timing
for the algebraic method
and the numerical integral plus root-finding
method in both configuration and Fourier space
(Figure~\ref{fig:timing}).  If we fit the polynomial coefficients to the
correlation function, our method
is faster than the naive Fourier space 
method by a 10--20$\times$; if we
generate the polynomial coefficients
from a Taylor expansion about the {\it Planck} 2018 cosmology, our  method is 40--80$\times$ faster than the Fourier space method.
Some of these gains are from working
in configuration space rather than in Fourier
space, where the integral is easier to evaluate. However, our algebraic
method is still  a factor of 2--4$\times$
faster than the naive method in configuration space.

The timing of the naive method has a slightly different
redshift dependence than that of the algebraic method.
The step features
in Figure~\ref{fig:timing} for the naive method
arise from discrete changes in the number
of steps needed to find $\Rnl$.
On the other hand, the timing
of the algebraic method slightly improves
with redshift, as at high $z$ we fit fewer points 
to determine the polynomial coefficients.

\section{Conclusions}

The nonlinear mass is the characteristic scale of halo formation,
defined as the scale on which $\sigma_R$, the rms of the density field inside a sphere 
of radius $R$,
reaches the linear threshold for spherical collapse, $\dc = 1.686$.
We present a method to accelerate computation
of the nonlinear mass by an order of magnitude
by fitting a polynomial to the correlation function
and evaluating $\sigma_R$ in configuration space.
Our method can be further accelerated
by a factor of 4 by using a Taylor
series about the {\it Planck} 2018 cosmology for the correlation function fitting coefficients.
We make our \textsf{python} implementation publicly
available at \url{https://github.com/akrolewski/NonlinearMassFaster}.

Overall, our method is sufficiently
accurate for future applications, with accuracy in the nonlinear
mass generally exceeding 1\% at $z < 6$
for a variety of cosmologies.
The accuracy is better at lower redshift,
where $\Rnl$ is larger and thus depends on the power
spectrum at lower $k$, where it is less
sensitive to cosmological parameters.

A fast and accurate method to compute nonlinear mass
will enable repeated calculations of the nonlinear mass. This would be necessary
in an MCMC chain
for cosmological analysis of a current or upcoming dataset
such as DES, DESI or LSST.
With this method one could simultaneously vary the nonlinear mass and the cosmology
in cosmological inference from the nonlinear regime,
potentially enabling more complex modeling
to extract cosmological information from smaller scales.

\section*{Acknowledgements}
AK thanks Antony Lewis for insight
into the high-$k$
linear power spectrum.
ZS thanks Hillary Child for bringing this problem to his attention and for useful input on this project. ZS is also grateful to both Lawrence Berkeley National Laboratory and the Berkeley Center for Cosmological Physics for hospitality during this work.




\bibliographystyle{mnras}
\bibliography{refs}








\bsp	
\label{lastpage}
\end{document}